\documentclass[aps,preprintnumbers,superscriptaddress,groupedaddress]{revtex4}
\usepackage{graphicx}  % needed for figures
\usepackage{dcolumn}   % needed for some tables
\usepackage{bm}        % for math
\usepackage{amssymb} \usepackage{amsmath}   % for math
\usepackage{color}
\definecolor{dgreen}{rgb}{0,0.70,0.30}
\definecolor{gold}{rgb}{0.85,.66,0}
\definecolor{purple}{rgb}{1.0,0.3,0.6}
\definecolor{red}{rgb}{1.0,0.0,0.0}

\def\sq[#1,#2]{\left[#1\,#2\right]}
\def\an[#1,#2]{\left\langle#1\,#2\right\rangle}
\def\spab[#1,#2,#3]{\left\langle#1|#2|#3\right]}

\begin{document}
\preprint{IPhT-T15/018, IHES/P/15/06}
\title{ The Equivalence Principle in a Quantum World}
\author{N.~E.~J~Bjerrum-Bohr}\email{bjbohr@nbi.dk}
\affiliation{Niels Bohr  International  Academy and  Discovery
Center,
The Niels Bohr Institute, Blegdamsvej 17,
DK-2100 Copenhagen \O, Denmark}
\author{John~F.~Donoghue}
\email{donoghue@physics.umass.edu}
\affiliation{Department of Physics-LGRT,
University of Massachusetts,
Amherst, MA 01003 USA}
\author{Basem~Kamal~El-Menoufi} \email{bmahmoud@physics.umass.edu}
\affiliation{Department of Physics-LGRT,
University of Massachusetts,
Amherst, MA 01003 USA}
\author{Barry~R.~Holstein} \email{holstein@physics.umass.edu}
\affiliation{Department of Physics-LGRT,
University of Massachusetts,
Amherst, MA 01003 USA}
\author{Ludovic~Plant\'e} \email{ludovic.plante@cea.fr}
\affiliation{Institut de physique th\'eorique,
 Universit\'e Paris Saclay, CEA, CNRS, F-91191 Gif-sur-Yvette}

\author{Pierre~Vanhove}\email{pierre.vanhove@cea.fr}
\affiliation{Institut de physique th\'eorique,
 Universit\'e Paris Saclay, CEA, CNRS, F-91191 Gif-sur-Yvette}

\affiliation{
Institut des Hautes {\'E}tudes Scientifiques
Bures-sur-Yvette, F-91440, France }

\date{\today}

\begin{abstract}
We show how modern methods can be applied to quantum gravity at low energy. We test how
quantum corrections challenge the classical framework behind the Equivalence Principle, for instance
through introduction of non-locality from quantum physics, embodied in the Uncertainty Principle.
When the energy is small we now have the tools to address this conflict explicitly.
Despite the violation of some classical concepts, the EP continues to provide the core of the
quantum gravity framework through the symmetry - general coordinate invariance -
that is used to organize the effective field theory.\\
\\ 
{\bf  Essay written for the Gravity Research Foundation 2015 Awards for Essays on
Gravitation. (Selected for Honorable Mention)}
\end{abstract}

%\pacs{04.60.-m, 04.62.+v, 04.80.Cc}
\maketitle

The Equivalence Principle (EP) in some form has been part of the study of gravity since the initial explorations of Galileo. Whether one is dropping objects of different
composition or rolling balls down inclines, the remarkable feature of gravity is that it acts on all objects universally. Once the force law was understood by
Newton, it was clear that the mass that enters $F=ma$, {\sl i.e.}, the inertial mass, had to be the same as the gravitational mass that appears in the gravitational
force.

However, it was at the hands of Einstein that the Equivalence Principle led us to a full theory of gravity. In 1907, Einstein had his ``happiest thought'', noting that an observer in free fall would not experience the effect of gravity. If the gravitational force can locally be removed by a choice of coordinates, then perhaps it should not be considered to be a force at all, but rather the manifestation of a curved spacetime. This reasoning had several consequences. If curved spacetime describes the manifestation of gravity, then particle motion in this setting would be described by geodesics. Also, the EP requires that gravity involves the total energy and momentum, not just the mass. If the free falling observer feels no gravity, then binding energies and the energy of photons also must gravitate - after all we have $E=mc^2$ and
 this leads to energy and momentum being the source of gravity. Even technical aspects such as the fact that
covariant differentiation in a curved spacetime must satisfy the metricity condition $D_\alpha g_{\mu\nu}(x)=0$ follow from the fact that the EP allows a locally flat coordinate system. It seems that all of the ingredients of General Relativity can be traced back in one way or another to the Equivalence Principle.

Of course, having been led to General Relativity by these conceptual considerations, we can look back and ask - what {\em exactly} is this Equivalence Principle? There are many answers and different versions of what one means by the Equivalence Principle~\cite{Will:2005va} - the Weak Equivalence Principle, the Strong Equivalence Principle, etc. In the end we cannot claim
one single formulation of the EP. However, certain operational consequences are clear. Tests
of the composition dependence of gravity must lead to null results. Such experiments have been performed and lead to stringent bounds on EP violation at the $10^{-12}$ level~\cite{tests}, with significant future improvements underway~\cite{future}. Massless particles in General Relativity must all move along null geodesics. General coordinate changes must be able to remove the effects of gravity in an infinitesimal local region.

In February 1927, sitting in his third floor office at Niels Bohr's institute, Werner Heisenberg had an equally fundamental thought - the Uncertainty Principle. Particles and events cannot be arbitrarily well localized. Low energy propagation is inherently spread out. The uncertainty relation $\Delta x\,\Delta p\geq \hbar/2$ is a fundamental property of quantum mechanics.  In quantum mechanics point particles are replaced by waves and the probabilistic interactions are obtained by computing transition amplitudes.

Obviously these two Principles clash.
If particles cannot be localized, how can a particle be contained in a locally flat region without gravity?
How can a geodesic even be defined if the particle samples more than one unique spacetime trajectory? Quantum physics require processes with the emission and absorption of quanta, and in such processes virtual massless particles such as the photon can propagate long distances, so what is the spacetime point relevant for the quantum interaction? Even worse, since gravity itself must be subjected to quantum physics and the
uncertainty principle, are there not fluctuations in spacetime itself? These are questions that make your head hurt!
But it seems quite plausible that the Equivalence Principle, given the quantum Uncertainty Principle, should be understood in a somewhat broader context.

The point of this essay is that we now have the tools to address and answer these questions through remarkable advances in both conceptual understanding and calculational capabilities.   The derived quantum effects are in ordinary situations exceptionally tiny but non-local, thus allowing us to probe the concepts of the Equivalence Principle in a quantum world.

\section{Gravitons and their fluctuations}

When we talk about ``spacetime fluctuating'', we really mean that the metric $g_{\mu\nu} (x)$ is the actor. Its dynamics is given by the Einstein-Hilbert Lagrangian
  \begin{equation}
    \mathcal L= {1\over 16\pi^2 G_N}\sqrt{-g} \, \mathcal R\,,
  \end{equation}
where $G_N$ is Newton's constant. The coordinates are arbitrary markers delineating spacetime, but the metric itself is the field variable. Classical gravitational waves, which we hope will soon be detected, are wavelike solutions to Einstein's equations for the metric. When quantum physics is considered, these waves become quantized, carrying energy $\hbar \omega$ and spin $\pm 2\hbar$. The quantization procedure was carried out by Feynman and DeWitt in the 1960's. By modern standards, the quantization itself is not problematic. However, subsequent investigations showed that the high-energy behavior of quantum corrections did not resemble that of renormalizable field theories. The resulting program searching for a consistent high-energy theory of quantum gravity continues to this day.

The most important point for our purposes is that the not-yet-known high-energy parts of quantum general relativity are not relevant for a discussion of the violation of the equivalence principle at ordinary energies. Indeed, it is Heisenberg who comes to our rescue in this regard. High energy fluctuations become
essentially local in spacetime. The uncertainty principle tells us that such quantum effects cannot propagate far,  since $\Delta x, ~\Delta t \sim \hbar/\Delta E$ and this is seen in explicit calculations. As we will describe more fully below, the one loop effects of pure gravity or of matter have a high energy divergence proportional to the local Lagrangian~\cite{'tHooft:1974bx}
\begin{equation}
\Delta {\cal L} = c_1 R_{\mu\nu}R^{\mu\nu}+ c_2R^2\,.
\label{gravdiv}
\end{equation}
This result is local and is fully invariant under general coordinate transformations, and satisfies the constraints of the Equivalence Principle.

On the contrary, the most non-local effects come from the low energy part of the theory. Here we know the particles involved and we know their gravitational interactions through the coupling of their stress-energy tensor to the metric. They are the usual massive constituents of matter and the massless fields mediating their interactions. At low-energy a massive object, of mass $M$, behaves as a classical source interacting gravitationally through the fluctuations of the metric of the spacetime fabric. At large distance the spacetime differs from flat space by post-Newtonian corrections organized into a power series in the Schwarzschild radius $2G_NM/c^2$ over the distance $r$
  \begin{equation}\label{e:gPPN}
g_{00} =1 -2 {G_NM\over c^2 r} +2 {G_N^2M^2\over c^4 r^2}+ \cdots
  \end{equation}
Quantum corrections to this metric (in harmonic gauge) arise through the nonlinear couplings of gravitons emitted by this sourse, yielding~\cite{BjerrumBohr:2002ks}
\begin{equation}\label{e:gQQ}
   \delta g_{00}= \hbar \, {62\over 15\pi} \, {G_N^2 M\over c^5r^3} +\cdots
 \end{equation}
These quantum effects are not sensitive to high-energy gravitational effects because they arise solely from the quantum uncertainty on the gravitational field
 surrounding the massive source.

The logic described above is that of Effective Field Theory (EFT). This technique provides a method for separating the known low-energy  physics from the high-energy physics which may be either unknown or dynamically irrelevant. Indeed in all of our fundamental theories we expect changes in the theory at the highest energy. EFT allows us to make reliable predictions using only the low-energy degrees of freedom and works even for theories that are deemed ``non-renormalizable''. Such theories can in fact be renormalized by including additional local terms in the Lagrangian, such as the one shown above in Eq.~\eqref{gravdiv} for general relativity~\cite{Gomis:1995jp}. But the divergences and other local effects are not predictions of the EFT - they arise from the high-energy side where the theory is not reliable. Rather in EFT we learn to focus on the predictions that follow from the reliable low energy end and such predictions often are non-local because of the Uncertainty Principle.

\section{Maintaining general covariance}

Reasoning from the Equivalence Principle has also led us to the idea of general covariance. A modern version of such reasoning might go as follows: From the EP we have determined that gravity must couple to total energy and momentum. In a relativistic theory, this implies that the energy-momentum tensor, or stress-energy tensor, $T_{\mu\nu}$ must serve as the source of gravity. We have learned from gauge theories that to get a given current as the source of a field we should gauge the corresponding global symmetry {\sl i.e., } make it local. Because energy and momentum conservation follows from invariance under space and time translations, to get $T_{\mu\nu}$ as a source we should make the space and time transformations local, leading to general covariance. Requiring such a symmetry leads to General Relativity.

Does this aspect of EP reasoning get damaged by quantum effects? Given all the bad press about the clash between quantum mechanics and general relativity, one might even expect this result. Indeed, there are more technical reasons for such worry. When solving the quantum theory, one must ``fix the gauge'', which for general relativity implies a constraint on the metric, selecting a subset of all possible coordinate choices. Through quantum loops, this might upset the coordinate invariance of the quantum results.

The most straightforward way to see that general covariance survives the quantization process is to use the background field method. Here one expands the metric around a background field which is arbitrary aside from being itself a solution to Einstein's equations.
Since we are interested in the gravitational scattering in a given classical background ({\it e.g.} a black hole or the Sun), it will be sufficient to compute the on-shell scattering amplitudes which lead to gauge-invariant observables without having to deal with the subtleties of defining an off-shell low-energy effective action.

\section{Scattering amplitudes}
Perturbative calculations in effective quantum gravity directly using Feynman diagrams is an extremely complicated and arduous task and is  far more involved and tedious than for gluon amplitudes in QCD. Luckily recent years have seen huge progress in the application of
on-shell techniques to compute amplitudes~\cite{Bern:2002kj}. Computing amplitudes in a traditional off-shell fashion creates a host of contributions that
in the end simply sum up to zero and this is avoided using  modern on-shell unitarity methods.
One very powerful character of this method is the reduction of quantum loop amplitudes to a combination of classical tree-level amplitudes.  Surprisingly, string theory teaches that tree-level gravity amplitudes can be simply computed from gauge theory amplitudes~\cite{Kawai:1985xq}. Even more remarkably some gravity amplitude are simply related to Abelian scalar QED amplitudes~\cite{Holstein:2006bh,Bjerrum-Bohr:2013bxa,Bjerrum-Bohr:2014lea}. The graviton is an ubiquitous particle, being strongly non-Abelian at high-energy, but has long range (infrared) properties closer to those of a photon~\cite{Weinberg:1965nx}.
Thus, recycling results from gauge theories using on-shell amplitudes and unitarity, we can directly generate results for gravitational loop amplitudes~\cite{Bern:2008qj,BjerrumBohr:2009zz}.  These simplifications allow an efficient~\cite{Bjerrum-Bohr:2013bxa} perturbative approach to the post-Newtonian corrections to the gravitational potential and their quantum corrections~\cite{Donoghue:1994dn,BjerrumBohr:2002kt}.
The modern unitarity approach allows an easy separation of the low-energy contributions and the bad high-energy behavior. There is  therefore no need to construct the full amplitude in order to isolate the long range effects. The results depend only on the low-energy data and their coupling to the classical background, and provide strong constraints on any fundamental formulation of quantum gravity.

As a consequence of the relation between gravity and gauge theory amplitudes, graviton scattering from a massive or a massless particle satisfies low-energy theorems similar to those of QED~\cite{Bjerrum-Bohr:2014lea}. All these simplifications offer the hope that even more elementary connections between gravity and gauge theories exist.

\section{Quantum Mechanics and the Equivalence Principle}

A century ago, with his new theory of general relativity Einstein derived the bending of light from a geodesic motion in the space curved by the Sun.
It is a classical result that the tree-level one-graviton exchange reproduces Newton's force~\cite{Veltman:1975vx}, so
one might well wonder about the result of evaluating the long range contribution due to two-graviton exchange, which is a one-loop effect, between a massless field of energy $\omega$ and the Sun of mass $M_\odot$.
In the low-energy limit, for small momentum transfer $q^2\simeq -\vec q^{\ 2}$, this scattering amplitude is given by~\cite{Bjerrum-Bohr:2014zsa}
\begin{eqnarray}
&&    i  \mathcal M_S^{\rm tree+1-loop}
 \simeq{1\over \hbar}\,  {(M_\odot\omega)^2\over 4 }
\,
 \Big[{\kappa^2\over {\vec q}^{\ 2}}+\kappa^4 {15\over
  512}{ M_\odot\over \sqrt{-{\vec q}^{\ 2}}} \cr
&+&\hbar \kappa^4
 {15\over
  512\pi^2}\,\log\left(-{\vec q}^{\ 2}\over M_\odot^2\right)
-\hbar\kappa^4\,{ bu^S \over64\pi^2} \, \log\left(-{\vec q}^{\ 2}\over \mu^2\right)
+ \hbar\kappa^4 {3\over128\pi^2}\, \log^2\left(-{\vec q}^{\ 2}\over \mu^2 \right)
-\kappa^4 \,  {M_\odot\omega\over 8\pi} {i\over {\vec q}^{\ 2}}\log\left(-{\vec q}^{\ 2}\over  M_\odot^2\right)\Big]\,.
\end{eqnarray}
where $bu^S$ a constant depending on the spin $S$ of the massless particle : $bu^0=3/40 $ and $bu^1=-161/120$.
The first line reproduces the leading gravitational potential as well as its  first post Newtonian correction to the gravitational potential in~\eqref{e:gPPN} computed by Einstein a century ago, and therefore satisfies the classical Equivalence Principle. This was not guaranteed but this is a consequence of the amplitude computation.
The second line gives the quantum corrections to the gravitational interaction between the massless field and the massive source.
Part of these corrections arise from the uncertainty in the fluctuations of the virtual massless particles and gravitons.
But there exists as well a dependence on the spin of the massless particle through the coefficient $bu^S$ due to the delocalized nature of a massless field, which induces tidal like effects.
Of course, this effect does not nessesarily lead to a violation of some versions of the Equivalence Principle, in that such effects are non-local in nature. However it emphasizes the feature
that our classical notions of gravity are scrutinized here and should be reconsidered in the new computational window.

We have observed some of the same type of effects relating to the bending of light in the case of a  very light charged particle in a loop~\cite{Basem}. Again the particle propagates a long distance, and this provides a non-local modification to the energy momentum tensor of the photon.
In this case, the helicity conserving amplitude of a photon scattering at small momentum transfer becomes
\begin{equation}
\mathcal{M}(++) =\mathcal{M}(--)\simeq \frac{(\kappa M_{\odot}\omega )^2}{2\hbar {\vec q}^{\ 2}} \left[1 - \frac{\beta^{(s,f)}}{e}\ln\left(\frac{{\vec q}^{\ 2}}{m^2}\right) \right] (1+\cos \theta)
\end{equation}
where $\beta^{(s,f)}$ is the QED beta function for either a scalar or fermion charged particle in the loop. The $\ln{\vec q}^{\ 2}$ term arises from the loop calculation. It corresponds to the non-local effect and, if translated into a na\"ive bending angle using the formalism of~\cite{Bjerrum-Bohr:2014zsa} would result in
\begin{align}
\theta \approx \frac{4G_N M_\odot }{c^2 b} + \frac{8  G_N M_\odot}{c^2 b} {\beta^{(s,f)} \over e}\left(\ln mb + \gamma_E - \ln 2\right) - \frac{4 \hbar G_N M_\odot}{ \omega^2b^3}  {\beta^{(s,f)} \over e}\ \ .
\end{align}
It appears that there are two effects here.
One is a non-universality, similar to our previous example, in that photons would have a different bending angle from other massless particles.
The other is an energy dependence of the bending angle which, however, could be a consequence of a too na\"ive use of the formalism of \cite{Bjerrum-Bohr:2014zsa}. It would be interesting to further explore this effect in a full quantum mechanical treatment. Both of the above examples clearly demonstrate the challenges to classical concepts like geodesics in a quantum world.

Classical descriptions of the Equivalence Principle use the notion of a {\em test particle} - a particle with no spatial extent which does not disturb the background geometry. Quantum mechanics and the uncertainty principle tell us that there is intrinsically no such thing as a test particle. The quantum effects described above are not optional and cannot be removed. Moreover, these are not Planck scale effects as one might na\"ively suppose in quantum gravity. They come
from the propagation of the lightest particles at low energy. It is also notable that we have now learned how to reliably calculate such properties even for the low-energy limit of quantum gravity.

\section{A quantum role for the Equivalence Principle}

The effective field theory formalism and the explicit calculations that we have described elucidate the role of the Equivalence Principle in a world
governed by quantum principles.
Some of the classical reasoning of the EP is seen to fail. There exist no point particles anymore because quantum gravitational corrections themselves involve propagation over long distances. These are not Planck scale effects. Because of the quantum uncertainty principle, massless particles at low energy sample regions that depend only on power-law suppression. The fact that this occurs only at low energy implies that the couplings are known from classical general relativity.  So overall, we conclude that reasoning with point particles is not valid.

One of the most important concepts to fail under quantum logic is the idea of geodesic motion. Quantum mechanics is not local enough to describe motion by a single trajectory in spacetime. This feature appears intuitively obvious when thinking about the structure of quantum physics. However, it is impressive that we have theoretical control over the combined quantum mechanics and general relativity theories when treated at low energies. The quantum corrections are not large, and for this reason they are calculable in perturbation theory at low energy.

However, we have seen that general coordinate invariance survives. Even if locations are now somewhat fuzzy because the metric fluctuates, the {\em symmetry} survives in the quantum theory and  it is this symmetry that allows us to organize the effective field theory systematically at low energies.
Even though some classical interpretations of the Equivalence Principle do not survive the addition of quantum mechanics, we have learned that nevertheless the symmetry implied by the Equivalence Principle still provides the primary organizing principle of low energy quantum gravity.

\end{document}